\pdfoutput=1
\documentclass[12pt]{article}

\usepackage[reqno]{amsmath}

\usepackage{mathrsfs}
\usepackage{amssymb}

\usepackage{geometry}
\usepackage{afterpage}
\usepackage[sort&compress,numbers,colon,merge]{natbib}
\usepackage{subcaption}
\usepackage{bbm}
\usepackage{epsfig}
\usepackage{array}
\usepackage{float}
\usepackage{color}
\usepackage{booktabs}

\usepackage{graphicx}

\usepackage{cancel}

\usepackage{wasysym}
\usepackage{url}
\usepackage{color}

\usepackage[colorlinks=true,linkcolor=red,citecolor=green,filecolor=magenta,urlcolor=cyan,bookmarks=true,bookmarksopen=true]{hyperref}

\def\gs{\mathrel{
   \rlap{\raise 0.511ex \hbox{$>$}}{\lower 0.511ex \hbox{$\sim$}}}}
\def\ls{\mathrel{
   \rlap{\raise 0.511ex \hbox{$<$}}{\lower 0.511ex \hbox{$\sim$}}}}

\newcommand{\be}{\begin{eqnarray}}
\newcommand{\ee}{\end{eqnarray}}

\newcommand{\hc}{\ensuremath{\mathrm{h.c.}}}

\newcommand{\tr}{\ensuremath{\mathrm{Tr}}}
\def\be{\begin{equation}}
\def\ee{\end{equation}}
\newcommand{\ba}{\begin{array}{c}}
\newcommand{\baz}{\begin{array}{cc}}
\newcommand{\bad}{\begin{array}{ccc}}
\newcommand{\bav}{\begin{array}{cccc}}
\newcommand{\baf}{\begin{array}{ccccc}}
\newcommand{\bena}{\begin{eqnarray}}
\newcommand{\eena}{\end{eqnarray}}
\newcommand{\bea}{\begin{equation} \begin{array}{c}}
\newcommand{\eea}{ \end{array} \end{equation}}
\newcommand{\ea}{\end{array}}

\newcommand{\Eqref}[1]{Eq.~(\ref{#1})}

\newcommand{\Figref}[1]{Fig.~\ref{#1}}
\newcommand{\Tabref}[1]{Tab.~\ref{#1}}

\usepackage{authblk}
\begin{document}

\begin{titlepage}
\begin{center}
{\Large\textbf{A Fresh Look at keV Sterile Neutrino Dark Matter from Frozen-In Scalars}}
\\[10mm]
{\large
Adisorn Adulpravitchai$^a$\footnote{\texttt{adisorn.adulpravitchai@gmail.com}}
and 
Michael A.~Schmidt$^{b}$\footnote{\texttt{m.schmidt@physics.usyd.edu.au}}}
\\[5mm]
{\small\textit{
$^a$Department of Physics, Faculty of Science, \\Chulalongkorn University,  Bangkok 10330, Thailand\\
$^b$ARC Centre of Excellence for Particle Physics at the Terascale,\\
School of Physics, The University of Sydney, NSW 2006, Australia
}
}
\end{center}

\vspace*{1.0cm}

\begin{abstract}
\noindent
Sterile neutrinos with a mass of a few keV can serve as cosmological warm dark matter. We study the production of keV sterile neutrinos in the early universe from the decay of a frozen-in scalar. Previous studies focused on heavy frozen-in scalars with masses above the Higgs mass leading to a hot spectrum for sterile neutrinos with masses below $8-10$ keV. Motivated by the recent hints for an X-ray line at 3.55 keV, we extend the analysis to lighter frozen-in scalars, which allow for a cooler spectrum. Below the electroweak phase transition, several qualitatively new channels start contributing. The most important ones are annihilation into electroweak vector bosons, particularly $W$-bosons as well as Higgs decay into pairs of frozen-in scalars when kinematically allowed.

\end{abstract}

\end{titlepage}

\setcounter{footnote}{0}

\section{Introduction}
Sterile Neutrinos are a well motivated minimal extension of the Standard Model (SM) of particle physics	~\cite{Abazajian:2012ys}. In particular sterile neutrinos with a keV-scale mass and tiny mixing with active neutrinos are able to explain the cosmological abundance of DM. The keV-scale of the sterile neutrinos can be explained in numerous ways, for instance, see \cite{Kusenko:2010ik,Lindner:2010wr,Adulpravitchai:2011rq,Merle:2011yv,Barry:2011fp,Merle:2013gea,Takahashi:2013eva}. In contrast to standard cold dark matter (CDM), they generally are
warmer and have a larger free-streaming horizon. Thus they can be
candidates for warm dark matter (WDM). As WDM candidates, they suppress structure at small scales and are able to solve the missing satellite problem~\cite{Kauffmann:1993gv,Klypin:1999uc,Moore:2005jj}. Additionally they might explain the velocities of pulsars~\cite{Kusenko:2006rh,*Kusenko:1997sp}.

As keV sterile neutrinos generally mix with the active neutrinos, they are produced non-thermally via oscillations~\cite{Barbieri:1990vx,Enqvist:1990ad}, which is called the Dodelson-Widrow (DW) mechanism~\cite{Dodelson:1993je} in the case of keV sterile neutrinos. 
The DW mechanism is excluded by observation~\cite{Canetti:2012kh,Canetti:2012vf}. However the bounds can be avoided, if there is a large enough primordial lepton asymmetry and sterile neutrinos are produced via resonant oscillations, the Shi-Fuller mechanism~\cite{Shi:1998km}. Further examples for production mechanisms are (a) the production from the decay of a scalar field which is in thermal equilibrium with the thermal plasma~\cite{Kusenko:2006rh,Petraki:2007gq,Frigerio:2014ifa}, (b) the production from inflaton decays \cite{Shaposhnikov:2006xi,Bezrukov:2009yw}, (c) the thermal production by new gauge interactions and subsequent dilution by production of entropy~\cite{Bezrukov:2009th,Nemevsek:2012cd} and (d) the production via the decay of a frozen-in scalar~\cite{Merle:2013wta}. 

We will focus on the production of keV sterile neutrinos from the decay of a frozen-in scalar~\cite{Merle:2013wta}. In this mechanism the sterile neutrino couples to a real scalar field and both couple extremely weakly to the thermal bath of SM particles. In addition to the active-sterile neutrino mixing, there is the coupling of the Higgs with the new scalar. The scalar is produced via freeze-in~\cite{Hall:2009bx} and subsequently decays to the keV sterile neutrino~\cite{Merle:2013wta}. The discussion in Ref.~\cite{Merle:2013wta} restricts the frozen-in scalar to be heavier than the SM Higgs. This however implies that keV sterile neutrinos with masses below $8-10$ keV form hot dark matter (HDM), erase too much structure at small scales, and thus are excluded.

This work is motivated by the recent hint for an X-ray line at approximately 3.55 keV by two analyses of the XMM-Newton data~\cite{Bulbul:2014sua,Boyarsky:2014jta}, which might be explained by sterile neutrino dark matter (DM) with a mass of approximately 7.1 keV~\cite{Ishida:2014dlp,Abazajian:2014gza,Bezrukov:2014nza,Chakraborty:2014tma,Haba:2014taa,Robinson:2014bma,Rodejohann:2014eka,Frigerio:2014ifa,Patra:2014pga}. Other explanations for the line are given in Ref.~\cite{Higaki:2014zua,Kong:2014gea,Frandsen:2014lfa,Baek:2014qwa,
Babu:2014pxa,Kolda:2014ppa,Queiroz:2014yna,Choi:2014tva,Liew:2014gia,Ko:2014xda,Dudas:2014ixa,Geng:2014zqa,Chiang:2014xra,Jaeckel:2014qea,Cicoli:2014bfa,Okada:2014zea,Lee:2014koa,Conlon:2014xsa,
Conlon:2014wna,Ishida:2014fra,Dubrovich:2014xaa,Nakayama:2014ova,
Bomark:2014yja,Nakayama:2014cza,Modak:2014vva,Cline:2014eaa,Baek:2014poa,Nakayama:2014rra,Chen:2014vna,Dutta:2014saa,Cline:2014kaa,Higaki:2014qua,Farzan:2014foo,Okada:2014oda,Faisel:2014gda,Boddy:2014qxa,Schutz:2014nka,Falkowski:2014sma}. 
 Irrespective whether the hints for the X-ray line persist~\cite{Boyarsky:2014ska,Boyarsky:2014paa} or disappear~\cite{Riemer-Sorensen:2014yda,Jeltema:2014qfa,Malyshev:2014xqa}, keV sterile neutrinos are excellent WDM candidates, including keV sterile neutrinos with a mass below 10 keV. Thus we are studying the production of a keV sterile neutrino from a frozen-in scalar without restricting the frozen-in scalar mass to be larger than the Higgs mass.

Above the electroweak (EW) phase transition, the only process contributing to the production of the frozen-in scalar is Higgs annihilation.\footnote{If a singlet condensate forms during inflation, there is an additional contribution to the production from the decay of the condensate.\cite{Enqvist:2014zqa}}
Frozen-in scalars lighter than the critical temperature of the EW phase transition lead to several new processes contributing to their production and consequently the DM abundance: SM fermion-antifermion annihilation, weak gauge boson annihilation, and Higgs decay to two frozen-in scalars if kinematically allowed. There is also the direct decay of the SM Higgs boson to a pair of keV sterile neutrinos. However it is negligible in the region of parameter space we study. The most important processes are Higgs annihilation and annihilation into EW vector bosons above $m_h/2$ and Higgs decay below $m_h/2$, when it becomes kinematically allowed. The fermion-antifermion annihilation is generally sub-dominant. 
Neutrino masses can be generated in different ways. We will firstly use an effective operator approach for neutrino mass and give one possible UV completion using the type-II seesaw mechanism~\cite{Magg:1980ut,Schechter:1980gr,Wetterich:1981bx,Lazarides:1980nt,Mohapatra:1980yp,Cheng:1980qt}.

The paper is organised as follows. In section~\ref{section-Model}, we introduce the model. In section~\ref{section-DMprod}, the dark matter production is explained. In section~\ref{section-FreeStream}, we discuss the free-streaming horizon of the dark matter in order to determine whether the keV sterile neutrino constitutes HDM, WDM, or CDM. In section~\ref{sec:UVcompletion} we give one possible UV completion of the effective operator for neutrino mass. Finally we conclude in section~\ref{section-Conclusion}. Technical details are collected in the appendices. Cross section and decay widths can be found in appendix~\ref{section-Cross}. Their thermal averages are given in appendix~\ref{section-Thermal}. A short summary of the treatment of effective degrees of freedom is given in appendix~\ref{section-effective}.

\section{Model} \label{section-Model}
Apart from the SM particle fermion content, we introduce one light SM singlet fermion $N$ and one scalar singlet $\phi$. For simplicity we introduce a discrete $Z_4$ version of lepton number\footnote{The breaking of the discrete symmetry $Z_4$ could lead to the formation of domain walls~\cite{Zeldovich:1974uw}, which alter the history of the universe. There are several different ways to avoid this problem. See for example ref.~\cite{Riva:2010jm,Larsson:1996sp,Dvali:1996zr,Dvali:1995cc,Preskill:1991kd}.  
The mechanism even works without imposing a $Z_4$ symmetry, but the following additional terms are introduced
\begin{equation*}
\mathcal{L}_{\cancel{Z_4}}= (\frac12 m_N^\prime N^2+\hc) + \rho_1 \phi + \rho_2 \phi^3 + \rho_3 H^\dagger H \phi\;.
\end{equation*}
Obviously the tadpole term of $\phi$ can be removed by shifting $\phi$. If the explicit mass term of $N$ is of the correct size and the coupling $\rho_3 H^\dagger H \phi$ is sufficiently small, the production of a keV sterile neutrino from a frozen-in scalar similarly works without the $Z_4$ symmetry.}  with the following transformation properties of the fields
\begin{align}
L &\rightarrow i L & 
E^c &\rightarrow -i E^c&
N &\rightarrow -i N &
H &\rightarrow H&
\phi &\rightarrow - \phi\;.
\end{align}
The most general Yukawa interactions in the lepton sector and the dimension 6 operator generating neutrino mass are given by 
\begin{equation}
-\mathcal{L}= y_{E} L H E^C + y_{LN} L H N    + \frac12 y_N \phi N^2 + \frac{y_\nu}{\Lambda^2}LLHH\phi+\hc\;.
\end{equation}
%
After all scalars obtain a vacuum expectation value (vev)
\begin{align}
H&=\begin{pmatrix}G^+\\ v+\frac{1}{\sqrt{2}} \left(h+ i G^0\right)\end{pmatrix}&
\phi&=v_\phi + \sigma\;,
\end{align}
with the Goldstone bosons $G^\pm$, $G^0$, the active neutrino mass matrix is given by
\begin{equation}
	m_\nu = y_\nu\, \frac{v_\phi v^2}{\Lambda^2} - \frac{y_{LN} y_{LN}^T }{y_N}\frac{v^2}{v_\phi}\;.
\end{equation}
We are interested in the region of parameter space with electroweak scale vevs $v\sim v_\phi$ and small couplings $y_{LN}^2\ll y_\nu y_N v_\phi^2/\Lambda^2 $, which results in a negligibly small seesaw contribution to the light neutrino mass.

The most general scalar potential is given by the scalar potential
\begin{align}
V=-\mu_H^2 H^\dagger H +\frac{\lambda_H}{6} (H^\dagger H)^2 
+\frac{\lambda_{H\phi}}{2} H^\dagger H \phi^2 
-\frac12\mu_\phi^2 \phi^2 + \frac{\lambda_\phi}{24} \phi^4
\end{align}
and the minimisation of the potential yields the vevs $v=\sqrt{3\mu_H^2/\lambda_H}$ and $v_\phi=\sqrt{6\mu_\phi^2/\lambda_\phi}$ in the limit of a small Higgs portal coupling $\lambda_{H\phi}$. At leading order there is no mixing between the different scalar fields. The scalar masses at leading order are given by 
\begin{equation}
m_h^2=2 \mu_H^2=\frac{2\lambda_H}{3}  v^2
\qquad\mathrm{and}\qquad 
m_\sigma^2 = 2\mu_\phi^2=\frac{\lambda_\phi}{3} v_\phi^2\;.
\end{equation}
The only relevant mixing for the production via freeze-in is the mixing between the Higgs doublet and the singlet $\phi$, which is given by 
\begin{equation}
\tan(2\gamma)= \frac{\lambda_{H\phi} }{\sqrt{2}} \frac{vv_\phi}{\mu_H^2-\mu_\phi^2}
=  \frac{3\sqrt{2}\lambda_{H\phi}vv_\phi}{2\lambda_H v^2-\lambda_\phi v_\phi^2}\;. \label{Higgs-S Mixing}
\end{equation}
The relevant couplings for the production of dark matter via freeze-in are the two Higgs portal interactions
\begin{align}
\Delta V &=
 \lambda_{H\phi} \frac{h^2 \sigma^2}{4} + \sqrt{2}\lambda_{H\phi}\,v \frac{h \sigma^2}{2}\;.
\end{align}
The decay of the sterile neutrino to an active neutrino and a photon is determined by the mixing with active neutrino flavour $\alpha$ via the small mixing angle
\begin{equation}
\theta_\alpha \approx  \frac{y_{LN,\alpha} v}{y_N v_\phi}\;.
\end{equation}
According to ~\cite{Bulbul:2014sua,Boyarsky:2014jta}, if we want to explain the 3.55 keV line from the decay of the sterile neutrino DM, the active-sterile mixing is constrained to be $\sum\limits_{\alpha} \sin^2(2\theta_\alpha) \simeq 7 \times 10^{-11}$.

\section{Dark Matter Production} \label{section-DMprod}
\begin{figure}[bt]\centering
\begin{subfigure}{0.49\linewidth}
\includegraphics[width=0.9\linewidth]{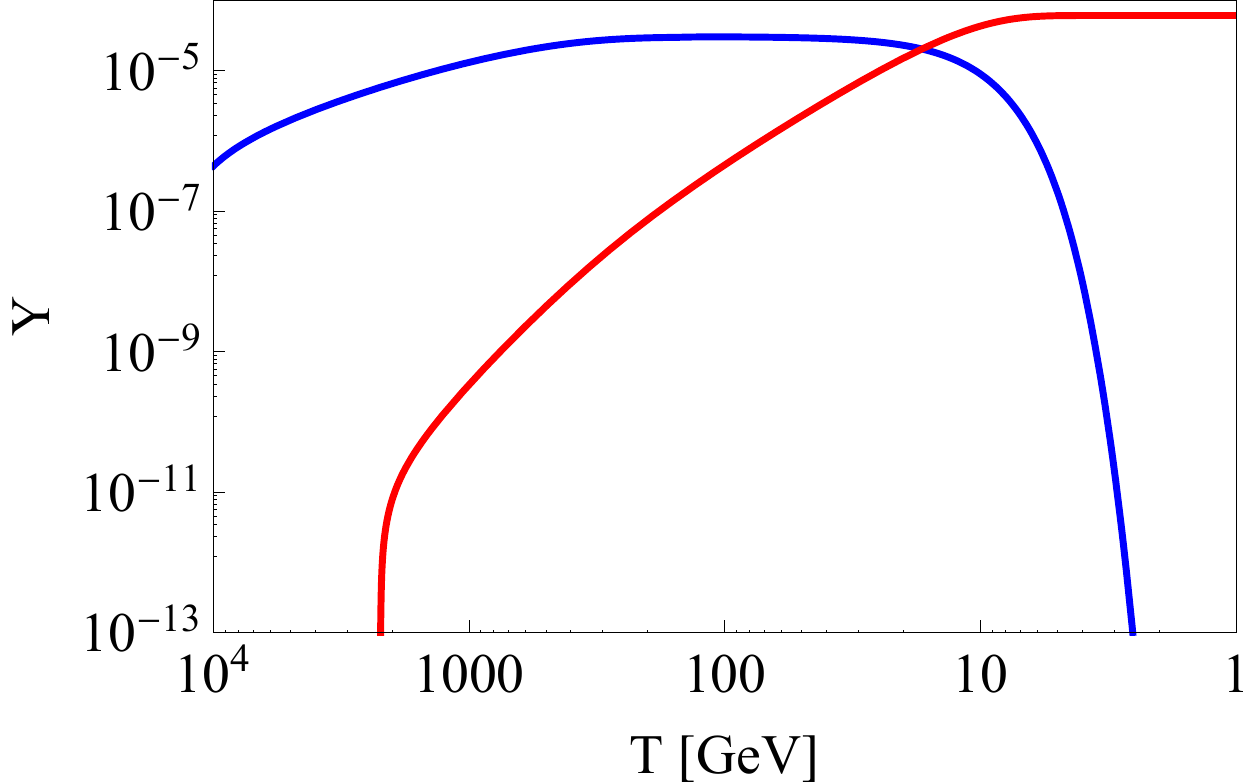}
\caption{$m_\sigma=500$ GeV}
\label{fig:DMprod180}
\end{subfigure}
\begin{subfigure}{0.49\linewidth}
\includegraphics[width=0.9\linewidth]{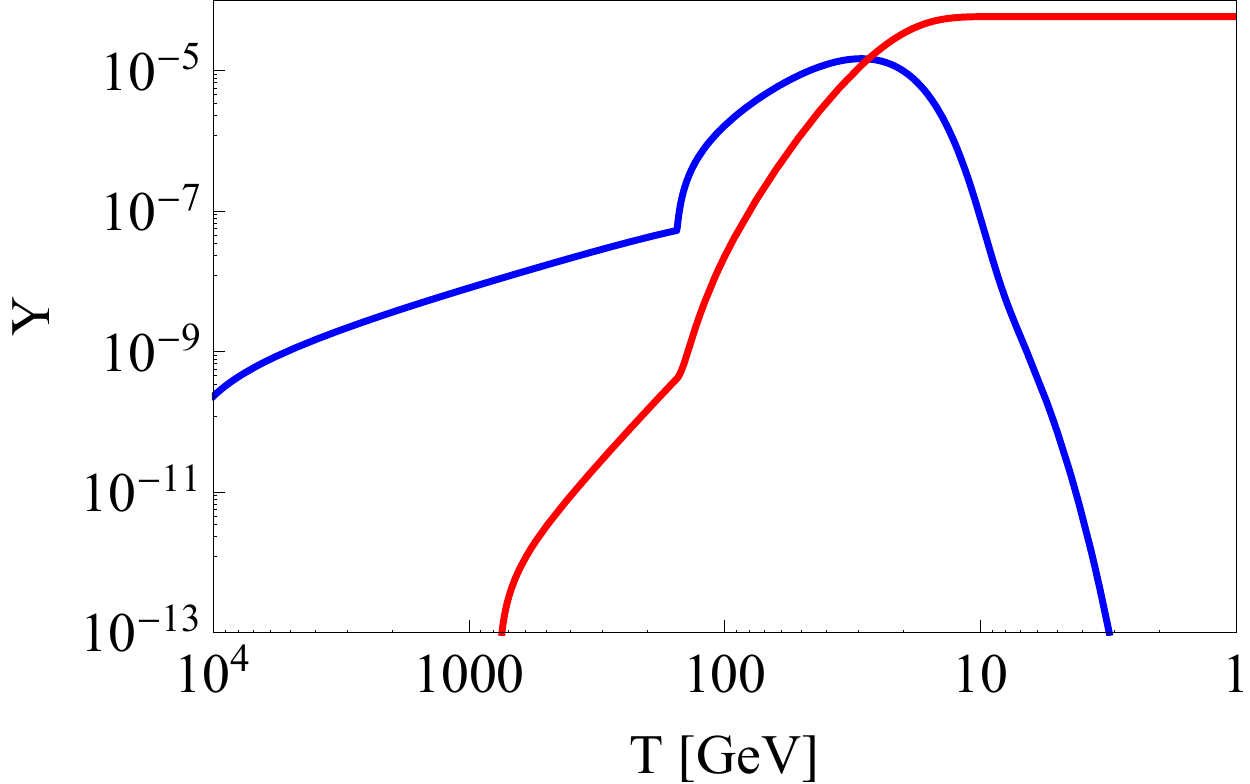}
\caption{$m_\sigma=60$ GeV}
\label{fig:DMprod60}
\end{subfigure}
\caption{Typical evolution of the abundances of the scalar $\sigma$ (shown in blue) and the keV sterile neutrino $N$ (shown in red). The sterile neutrino mass is fixed to $m_N=7.1$ keV and $\lambda_\phi=0.5$.}
\label{fig:DMprod}
\end{figure}

We consider the sterile neutrino production in the early universe. If the keV sterile neutrino gets into thermal equilibrium with SM particles, it will be overproduced and will overclose the universe after freeze-out. Therefore usually a keV sterile neutrino is very weakly coupled to the thermal bath and it is assumed that the initial abundance is zero. We concentrate on the freeze-in mechanism as main production mechanism, which has been first discussed in \cite{Merle:2013wta}.   In the freeze-in mechanism, keV sterile neutrinos are produced in two steps. First a feebly-coupled scalar field, $\sigma$, is produced via a tiny Higgs portal coupling, $\lambda_{H \phi}$, which has to be small enough such that $\sigma$ is always out of thermal equilibrium. This scalar subsequently decays into keV sterile neutrinos. The authors of Ref.~\cite{Merle:2013wta} consider scalars with a mass heavier than the Higgs mass. In this region of parameter space the main process is Higgs annihilation ($hh\leftrightarrow \sigma\sigma$). 
See \Figref{fig:DMprod180} for a typical evolution of the abundances of the keV sterile neutrino (red) and the scalar $\sigma$ (blue) for a frozen-in scalar with a mass larger than the mass of the Higgs. 
As it can be seen in \cite{Merle:2013wta} and we explain in more detail in the next section, it turns out that heavy scalars lead to larger free-streaming scales for light keV sterile neutrinos, like a $7.1$ keV sterile neutrino. Thus it is interesting to consider scalars lighter than the Higgs and consequently below the EW phase transition. In addition to Higgs annihilation, there are the following additional processes:
\begin{itemize}
\item annihilation of vector bosons: $VV\leftrightarrow \sigma\sigma$ with $V=W,Z$,
\item annihilation of SM fermions: $\bar f f \leftrightarrow \sigma\sigma$ 
\item Higgs decay to pairs of the scalar $\sigma$ as well as pairs of keV sterile neutrinos directly.
\end{itemize}
A typical evolution for a frozen-in scalar with mass $m_\sigma=60$ GeV is shown in \Figref{fig:DMprod60}. Note that the production of the keV sterile neutrino is dominated by Higgs decay and very weakly depends on the physics above the EW phase transition. 
The abundances $Y_{\sigma,N}\equiv n_{\sigma,N}/s$ normalised to the entropy density $s$ are described by the following Boltzmann equations:
\begin{align}
\frac{d Y_{\sigma}}{d T} =&  \frac{d Y_{\sigma}^{\rm A}}{d T} + \frac{d Y_{\sigma}^{\rm D}}{d T} +  \frac{d Y_{\sigma}^{\rm HD}}{d T}
&
\frac{d Y_{N}}{d T} =& \frac{d Y_{N}^{\rm D}}{d T} +\frac{d Y_{N}^{\rm HD}}{d T}
\end{align}
where the superscripts A, D, and HD denote annihilation, decay and Higgs decay terms, respectively. Before the SM particles get out of thermal equilibrium, the different terms are given by
\begin{subequations}
\begin{align}
\frac{d Y_{\sigma}^{\rm A}}{d T} =&   \sqrt{\frac{\pi}{45 G_N}} \sqrt{g_*(T)}  \sum_{i=h,W,Z,t,b,c,\tau} \langle \sigma v(\sigma \sigma \rightarrow ii) \rangle \big(Y_\sigma (T)^2- Y_\sigma^{eq}(T)^2\big) 
\\
\frac{d Y_{\sigma}^{\rm D}}{d T}  =&  -\frac{1}{2}\frac{d Y_{N}^{\rm D}}{d T}
\\
 \frac{d Y_{N}^{\rm D}}{d T}  =& -\sqrt{\frac{45}{\pi^3 G_N}} \frac{1}{T^3} \frac{1}{\sqrt{g_{eff}(T)}} \langle \Gamma(\sigma \rightarrow N N)\rangle \left(Y_\sigma (T)-\left(\frac{Y_{N}(T)}{Y_{N}^{eq}(T)}\right)^2 Y_\sigma^{eq}(T)\right) \\
\frac{d Y_{\sigma}^{\rm HD}}{d T}  =& -\sqrt{\frac{45}{\pi^3 G_N}} \frac{1}{T^3} \frac{1}{\sqrt{g_{eff}(T)}} \left\langle \Gamma(H \rightarrow \sigma \sigma) \right\rangle \left(1-\left(\frac{Y_\sigma(T)}{Y_\sigma^{eq}(T)}\right)^2\right) Y_h^{eq}(T) 
\\
\frac{d Y_{N}^{\rm HD}}{d T}  =& -\sqrt{\frac{45}{\pi^3 G_N}} \frac{1}{T^3} \frac{1}{\sqrt{g_{eff}(T)}} \langle \Gamma(H \rightarrow N N)\rangle \left(1-\left(\frac{Y_N(T)}{Y_N^{eq}(T)}\right)^2\right) Y_h^{eq}(T)  
\;.
 \end{align}
\end{subequations}
 $Y^{eq}_X(T)=n^{eq}_X(T)/s(T)$ denotes the equilibrium abundance, $G_N$ Newtons constant, $g_{eff}(T)$ the effective degrees of freedom at temperature $T$, and $g_*(T)$ is defined in the usual way. The precise definitions of the thermally averaged cross sections and decay widths as well as $g_{eff}$ and $g_*$ are collected in the appendices.
\begin{table}[bt]\centering
\begin{tabular}{lcccccc}\toprule
$m_{\sigma}$ [GeV] 
& 30 & 60 & 65 & 100 & 170 & 500\\
 $\lambda_{H\phi}$ [$10^{-9}$] 
& 3.65 & 6.26 & 62.6 & 75.0  & 116 & 274\\
$\Omega_N h^2$ 
& 0.1174 & 0.1199 &  0.1188 & 0.1175 & 0.1218 &0.1239\\
$T_{in}$ [GeV] 
& 17.8& 16.3 & 18.1 & 16.4& 13.8 & 8.77\\
$r_{FS}$ [Mpc] 
&0.018 & 0.035  &0.034 &0.054 &0.10 & 0.40 \\
\bottomrule
\end{tabular}
\vspace{1ex}

\begin{minipage}{0.85\linewidth}
\caption{Benchmark points. We fix $\lambda_\phi=0.5$ and the keV sterile neutrino mass $m_N=7.1$ keV. $T_{in}$ denotes the temperature when 80\% of the keV sterile neutrino DM are produced and $r_{FS}$ denotes the free-streaming horizon scale, which is discussed in sec.~\ref{section-FreeStream}. }
\label{tab:paramChoice}
\end{minipage}
\end{table}
\begin{figure}[p!]\centering
\begin{subfigure}{0.49\linewidth}
\includegraphics[width=\linewidth]{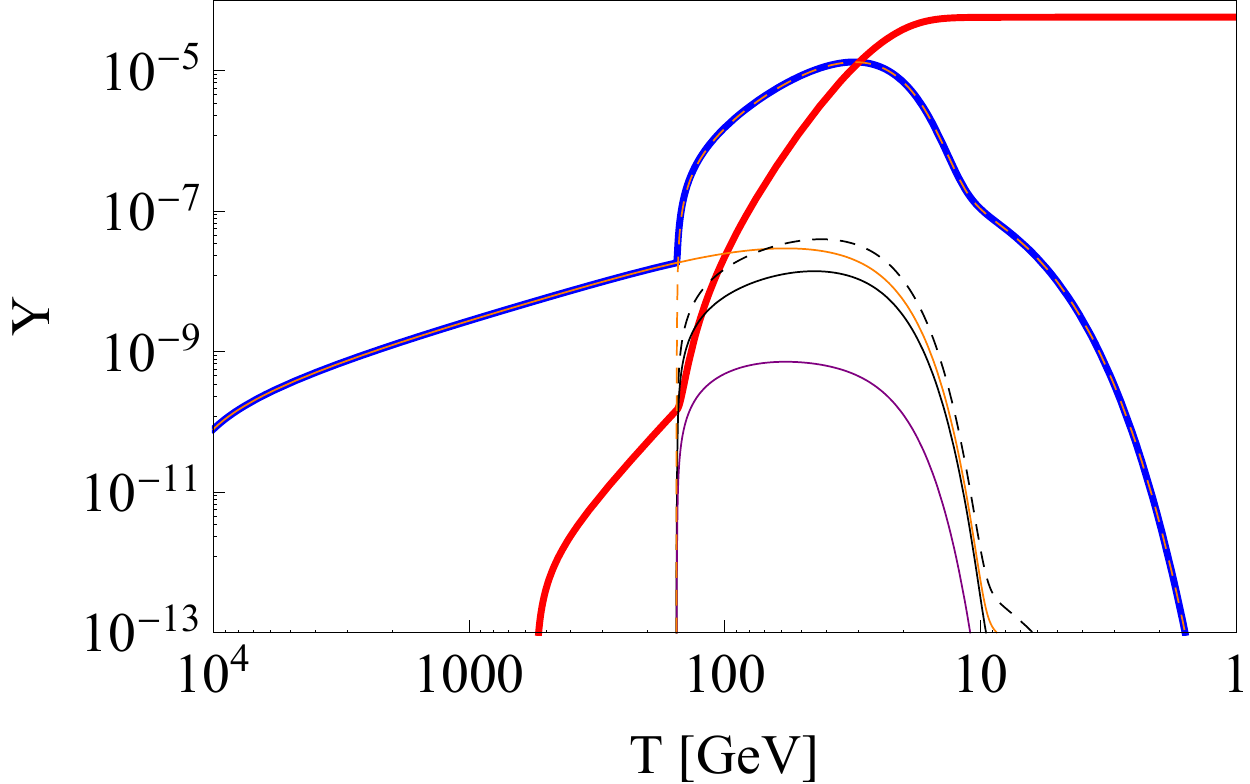}
\caption{$m_\sigma=30$GeV}
\label{fig:DMprod30b}
\end{subfigure}
\begin{subfigure}{0.49\linewidth}
\includegraphics[width=\linewidth]{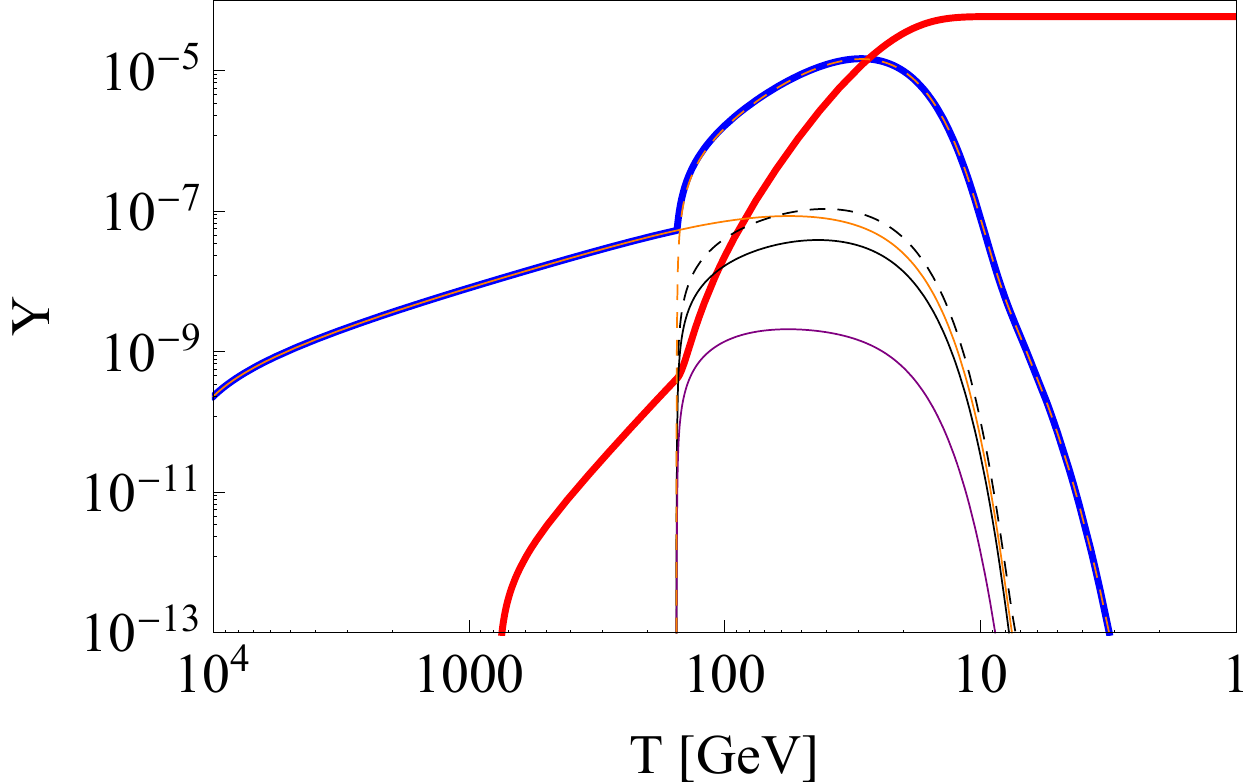}
\caption{$m_\sigma=60$GeV}
\label{fig:DMprod60b}
\end{subfigure}

\begin{subfigure}{0.49\linewidth}
\includegraphics[width=\linewidth]{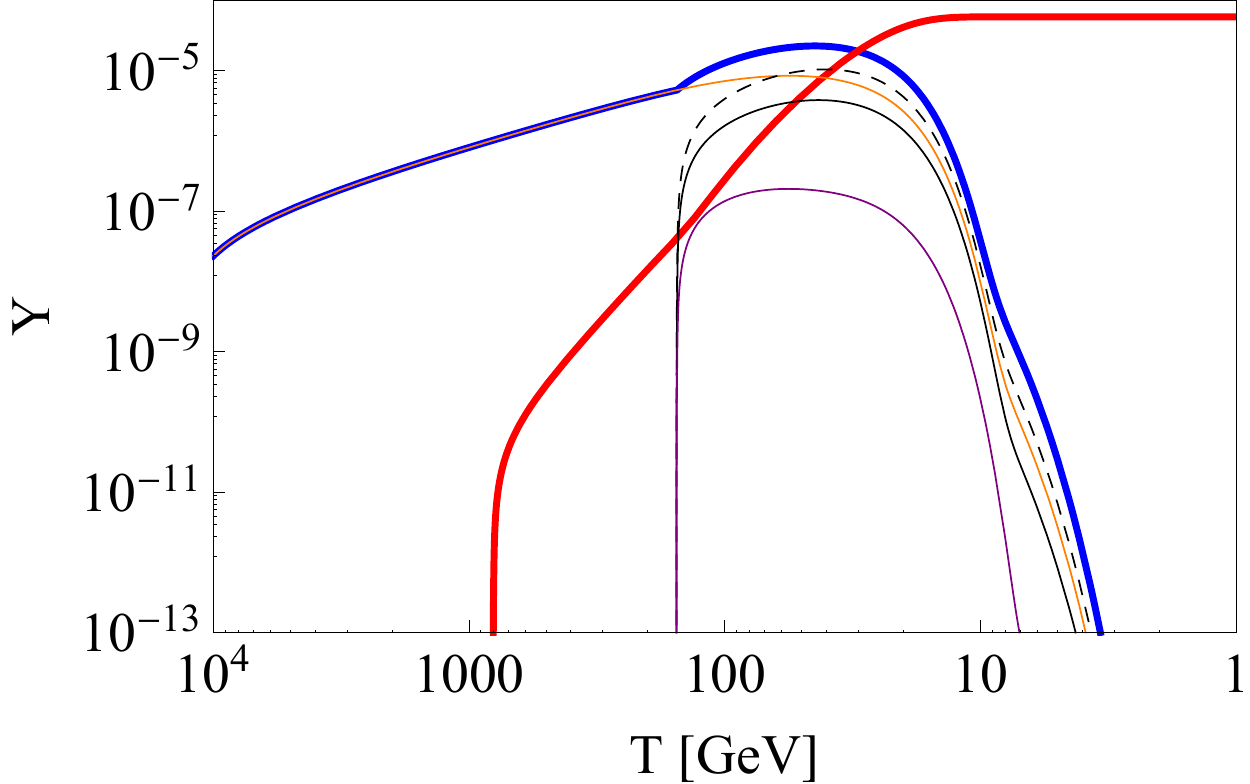}
\caption{$m_\sigma=65$GeV}
\label{fig:DMprod65b}
\end{subfigure}
\begin{subfigure}{0.49\linewidth}
\includegraphics[width=\linewidth]{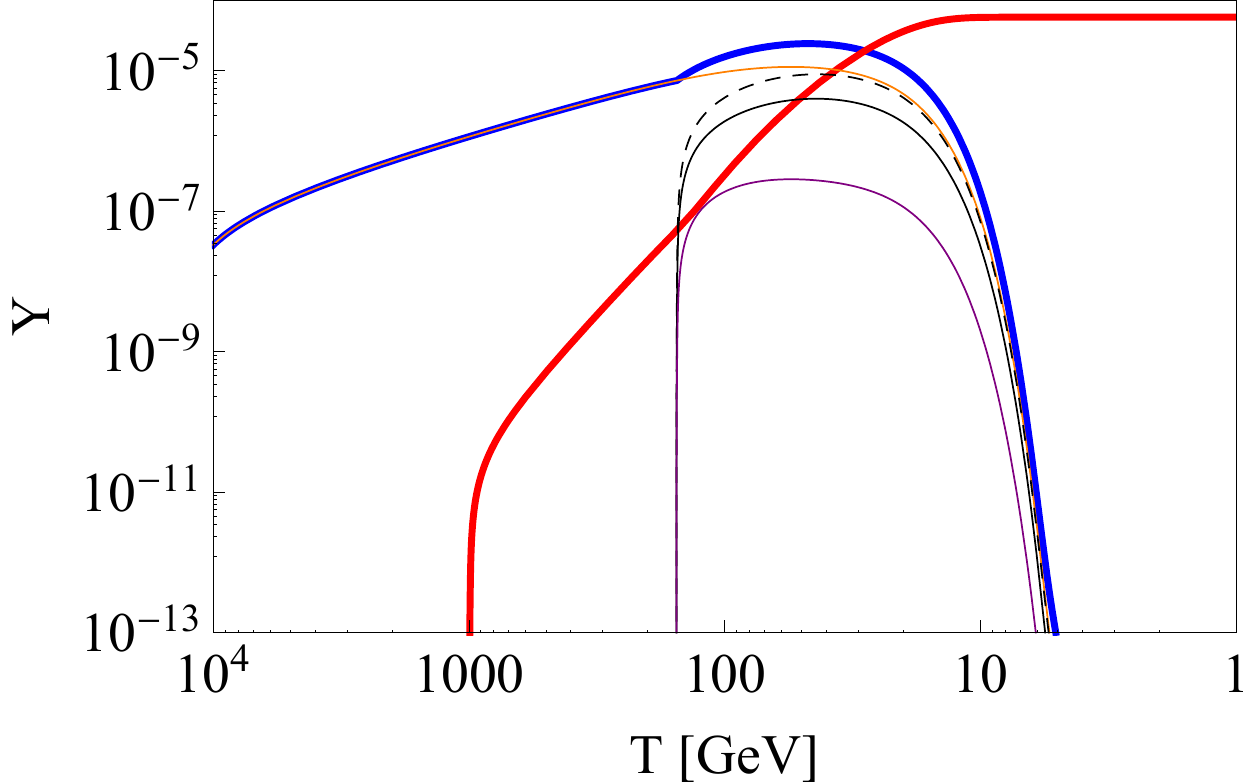}
\caption{$m_\sigma=100$GeV}
\label{fig:DMprod100b}
\end{subfigure}

\begin{subfigure}{0.49\linewidth}
\includegraphics[width=\linewidth]{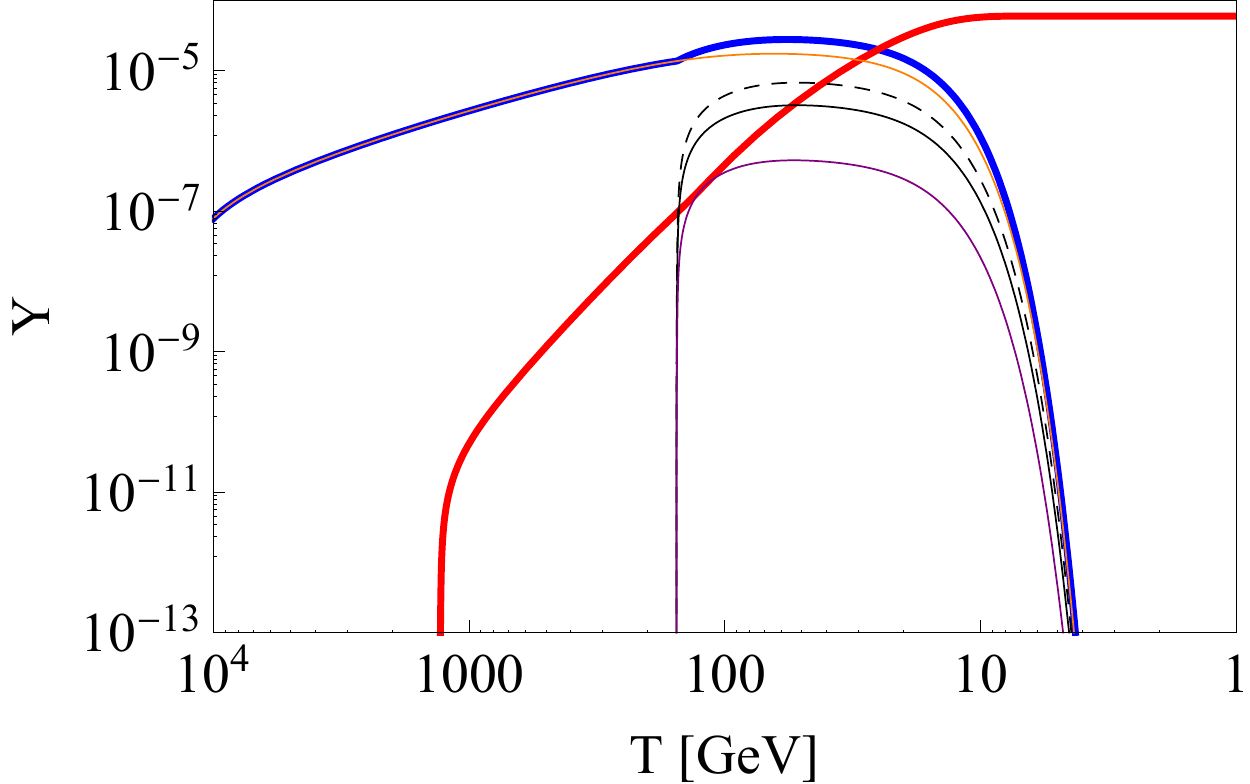}
\caption{$m_\sigma=170$GeV}
\label{fig:DMprod170b}
\end{subfigure}
\begin{subfigure}{0.49\linewidth}
\includegraphics[width=\linewidth]{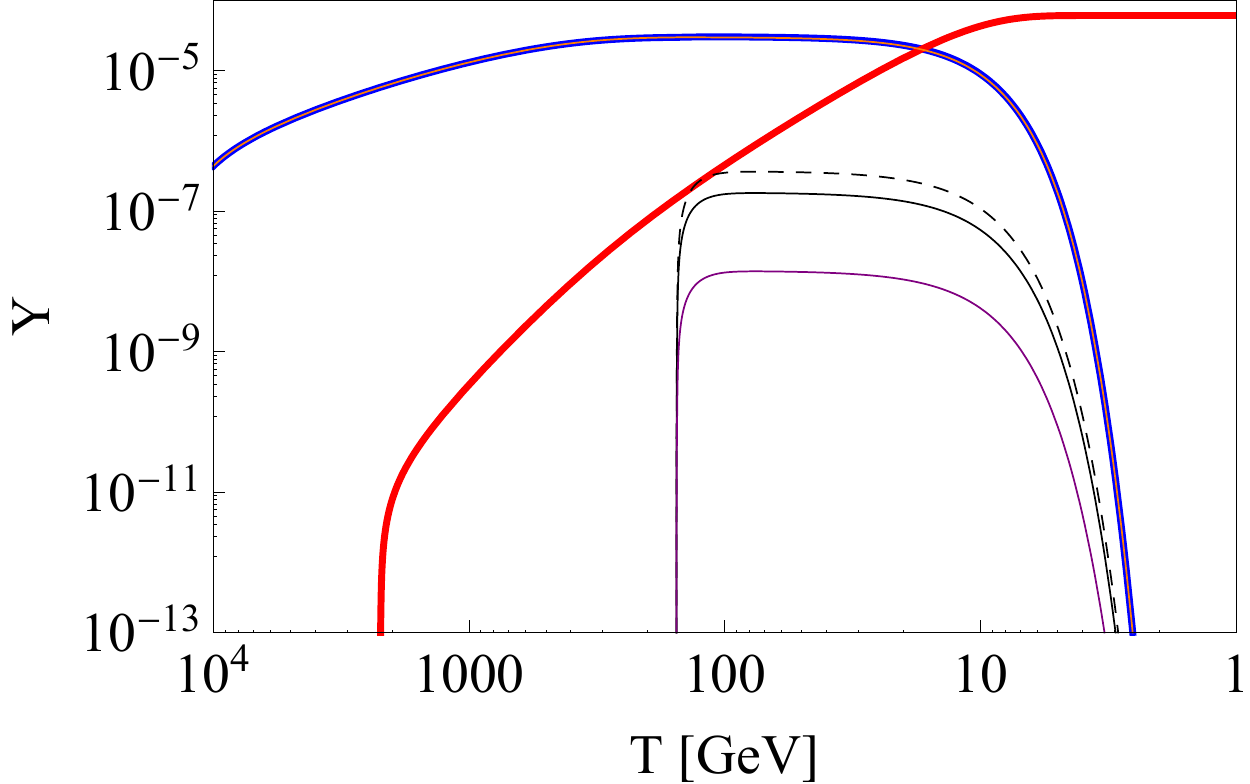}
\caption{$m_\sigma=500$GeV}
\label{fig:DMprod500b}
\end{subfigure}
\
\caption{DM production for six benchmark points. The thick red (blue) line indicates the abundance of the keV sterile neutrino (scalar $\sigma$). The orange solid (dashed) line describes the contributions of Higgs annihilation (decay) to the abundance of $\sigma$. The annihilation of $Z$-bosons ($W$-bosons) is described by the black solid (dashed) line. The contribution of $t\bar t$ annihilations are shown in purple.}
\label{fig:DMprod2}
\end{figure}
We show the different contributions for six different benchmark points. We fix the keV sterile neutrino mass to $m_N=7.1$ keV and the scalar self-coupling $\lambda_\phi=0.5$, which fixes the vev of the scalar singlet $\sigma$. We choose the Higgs portal coupling $\lambda_{H\phi}$ such that the observed DM abundance of $\Omega_{DM}=0.1199\pm0.0027$~\cite{Ade:2013lta} is obtained at $2\sigma$. We also include the contribution from the DW mechanism using the approximate formula~\cite{Kusenko:2009up}
\begin{equation}
\Omega_{N,DW}h^2\approx 0.2\times \frac{\sum_\alpha\sin^2\theta_\alpha}{3\times 10^{-9}}\left(\frac{m_N}{3 \mathrm{keV}}\right)^{1.8}
\end{equation}
fixing $\sum_\alpha \sin^2(2\theta_\alpha)=7\times 10^{-11}$.
All parameter choices are collected in \Tabref{tab:paramChoice}. The scalar vev, $v_\phi$, varies between 73 GeV and 1.3 TeV for our chosen benchmark points. The evolution of the abundances and the different contributions are shown in \Figref{fig:DMprod2} for the six benchmark points.

Similarly to \Figref{fig:DMprod}, the blue solid line shows the evolution of the abundance of the scalar $\sigma$ and the red solid line, the evolution of the abundance of the keV sterile neutrino. The orange solid (dashed) line describes the contributions of Higgs annihilation (decay) to the abundance of $\sigma$. The annihilation of $Z$-bosons ($W$-bosons) is described by the black solid (dashed) line. The contribution of $t\bar t$ annihilations are shown in purple. The annihilation into light fermions is generally sub-dominant compared to $t\bar t$ annihilation. (See e.g.~\cite{Frigerio:2011in,Babu:2014pxa}). Note that the contribution from the Higgs decay to pairs of sterile neutrinos is negligible for our chosen parameters. We do not show its contribution in the plots. Higgs annihilation is the only process above the EW phase transition, because the other processes are proportional to the EW vev and are absent above the EW phase transition. In our numerical analysis we simply set the corresponding annihilation cross section to zero above the EW phase transition. A proper treatment requires the inclusion of thermal corrections to properly treat the temperature dependence during the EW phase transition. Below the EW phase transition we have to distinguish between frozen-in scalars with masses $m_\sigma>m_h/2$ and $m_\sigma<m_h/2$: for $m_\sigma>m_h/2$ there is a sizeable contribution from EW gauge boson annihilations, particularly annihilation of $W$-bosons, besides Higgs annihilation, as it can be seen in the plots for $m_\sigma=65,100$ GeV, but for $m_\sigma<m_h/2$, the most important process is Higgs decay (See the plots for $m_\sigma=30$ GeV and $m_\sigma=60$ GeV.). As the coupling $\lambda_{H\phi}$ is extremely small, the contribution to the invisible decay width of the Higgs is negligibly small.

\section{Free-Streaming Horizon} \label{section-FreeStream}
The free-streaming horizon can be used as the indicator whether the keV sterile neutrinos are HDM, WDM, or CDM. It is the comoving mean distance which a collision-less gravitationally unbound particle travels 
\begin{equation}
r_{\rm FS} = \int^{t_0}_{t_{\rm in}} \frac{\langle v(t) \rangle}{a(t)}dt \;,
\end{equation}
where $t_0$ denotes time today, $t_{\rm in}$ is its production time, $\langle v(t) \rangle$ its mean velocity at time $t$, and $a(t)$ the scale factor at time $t$. 

In our discussion we will neglect the DW contribution and concentrate on the main production mechanism via decays of frozen-in scalars. The DW contribution generally leads to a larger free-streaming horizon scale, since it is hotter than the contribution from the frozen-in scalar. For the parameters that we have chosen, the contribution of the DW mechanism represents less than 5\% of the cosmological abundance, which is consistent with all known constraints. 
A detailed study of the free-streaming horizon scale is beyond the scope of this paper.

Guided by our numerical results in \Figref{fig:DMprod2}, we estimate the production time $t_{\rm in}$\footnote{The relation between the production time $t_{\rm in}$ and the production temperature $T_{\rm in}$ is given by $t_{\rm in} \simeq \left(1.5 g_{*,S}^{-1/4}\right)^2 \left(\frac{{\rm MeV}}{T_{\rm in}}\right)^2\mathrm{sec}$.} of the keV sterile neutrinos to be the time, when 80\% have been produced.\footnote{This differs from the approach of ref.~\cite{Merle:2013wta}, where the production time was estimated as $t_{\rm in}=t_{\rm prod, \sigma} + \tau$ where $t_{\rm prod,\sigma}$ ($\tau$) is the production (decay) time of $\sigma$. In ref.~\cite{Merle:2013wta} the production time $t_{\rm prod,\sigma}$ was estimated to be the time when $\sigma$ becomes non-relativistic which is correct above the Higgs mass, but breaks down for light scalars $\sigma$. Our definition agrees with the one used in ref.~\cite{Merle:2013wta} for $m_\sigma \gtrsim 100$ GeV.}

We follow the discussion of the free-streaming horizon in~\cite{Merle:2013wta}. See Ref.~\cite{Hasenkamp:2012ii} for an earlier discussion applied to gravitino DM. We assume an instantaneous transition between the relativistic and the non-relativistic regimes of the keV sterile neutrino, i.e. $\langle v(t) \rangle \simeq 1$ for $t<t_{\rm nr}$ and $\langle v(t) \rangle \simeq \frac{\langle p(t) \rangle}{m_N}$ for $t \geq t_{\rm nr}$, where $t_{\rm nr}$ is the time when the particle becomes non-relativistic, that is $\langle p(t_{\rm nr}) \rangle=m_N$. 

Our numerical results in \Figref{fig:DMprod2} show that the frozen-in scalar with masses $m_\sigma\gtrsim 30$ GeV mostly decays when it is non-relativistic. If the frozen-in scalar is lighter, a significant fraction will already decay when they are relativistic. We will focus on frozen-in scalar with masses $m_\sigma\gtrsim 30$ GeV and thus can safely assume that it decays non-relativistically. The rest of the discussion follows ref.~\cite{Merle:2013wta}.
The distribution of the DM produced from a non-relativistic parent $\sigma$ is given by~\cite{Kamada:2013sh,Aoyama:2011ba,Strigari:2006jf,Kaplinghat:2005sy}
\begin{equation}
f(p,t) = \frac{\beta}{p/T_{\rm DM}} e^{-p^2/T_{\rm DM}^2},
\end{equation}
where $\beta$ is a normalisation factor and $T_{\rm DM}=p_{cm}a(t_d)/a(t)$ is the DM temperature. 
Using the DM momentum in the centre-of-mass frame, 
$$p_{cm}=\frac{\sqrt{m_\sigma^2-m_N^2}}{2}\simeq \frac{m_\sigma}{2}$$ and the decay time $t_d \equiv t_{\rm in}$, 
\begin{equation}
T_{\rm DM}(t) \simeq \frac{m_\sigma}{2}\frac{a(t_{\rm in})}{a(t)} \;.
\end{equation}

The average momentum $\langle p(t)\rangle$ can be calculated to be 
\begin{equation}
 \langle p(t)\rangle = \frac{\int d^3p p f(p,t)}{\int d^3 p f(p,t)} = \frac{\int^{\infty}_0 dp p^2 e^{-p^2/T^2_{\rm DM}}}{\int^{\infty}_0 dp p e^{-p^2/T^2_{\rm DM}}} = \frac{\sqrt{\pi}}{2} T_{\rm DM}  \;. \label{p-average}
\end{equation}

In the radiation dominated era, the scale factor $a\propto t^{1/2}$. Thus from \Eqref{p-average} and $\langle p(t_{\rm nr}) \rangle=m_N$, the time when $N$ becomes non-relativistic is given by
\begin{equation}
t_{\rm nr} = \frac{\pi}{16} \frac{m_\sigma^2}{m_N^2} t_{\rm in} 
\end{equation}
\begin{figure}[tb!]\centering
\includegraphics[width=0.7\linewidth]{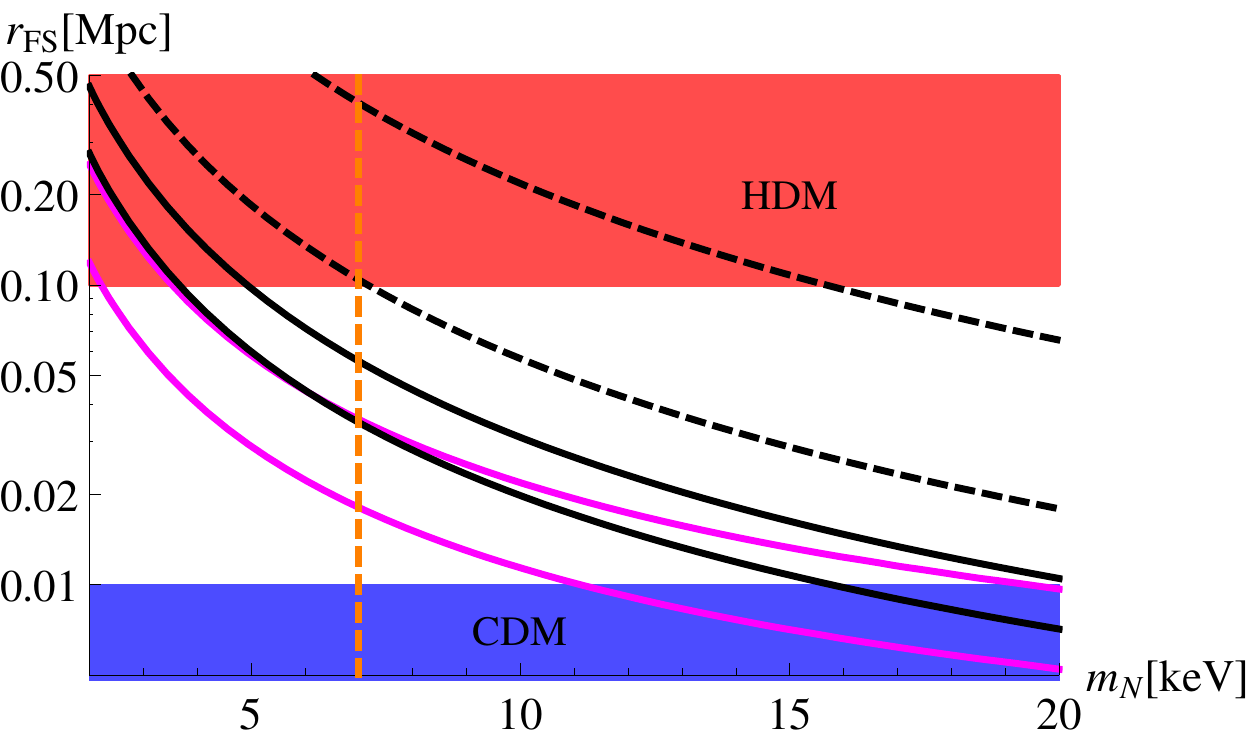}
\caption{Free-streaming horizon vs. keV sterile neutrino mass. The regions of HDM and CDM are marked red and blue, respectively. The black and magenta lines show the free-streaming scale $r_{FS}$ for different masses of the frozen-in scalar. We assume a vanishing active-sterile mixing, $\theta=0$. From the lowest to the highest line the masses of the frozen-in scalar are $m_\sigma=30,60$ GeV (magenta solid lines), $m_\sigma=65,100$ GeV (black solid lines) and $m_\sigma=170,500$ GeV (black dashed lines). The vertical orange dashed line indicates $m_N=7.1$ keV. }
\label{FreeStream}
\end{figure}%
\begin{figure}[tb!]\centering
\includegraphics[width=0.7\linewidth]{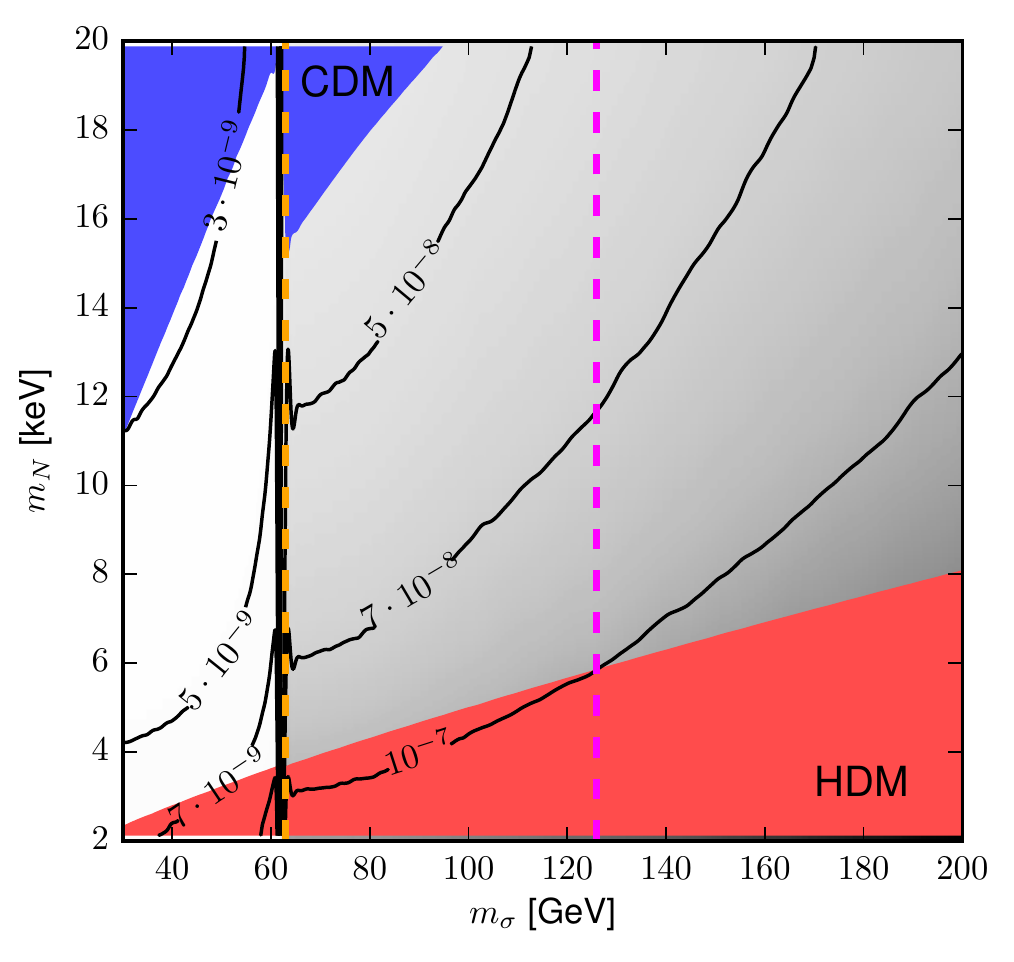}
\caption{Contour plot of $\lambda_{H\phi}$ in the plane of $m_N$ vs $m_\sigma$ requiring that the abundance of the keV sterile neutrino can account for the DM abundance at $1\sigma$, i.e. $0.1172<\Omega_N h^2<0.1226$~\cite{Ade:2013lta}. The (blue) region above ($r_{FS} \leq 0.01$ Mpc.) is the area of CDM while the (red) region below ($r_{FS} \geq 0.1$ Mpc.) is the area of HDM. The area in between ($0.01 {\rm Mpc} \leq r_{\rm FS} \leq 0.1 {\rm Mpc}$) is the WDM region. The magenta line indicates the Higgs mass $m_h$ and the orange line is at $m_h/2$. Below the orange line Higgs decays dominate the production of $\sigma$.}
\label{fig:scan}
\end{figure}
In our case DM becomes non-relativistic before the time of matter-radiation equality, $t_{\rm eq}=1.9 \times 10^{11} {\rm s}$. Thus the free-streaming horizon can be calculated as
\begin{eqnarray}
r_{\rm FS} &=& \int_{t_{\rm in}}^{t_0} \frac{\langle v(t) \rangle}{a(t)} dt = \int_{t_{\rm in}}^{t_{\rm nr}} \frac{1}{a(t)} dt+\int_{t_{\rm nr}}^{t_{\rm eq}} \frac{\langle v(t) \rangle}{a(t)} dt+\int_{t_{\rm eq}}^{t_0} \frac{\langle v(t) \rangle}{a(t)} dt\\
&\simeq& \frac{2\sqrt{t_{eq}t_{nr}}}{a_{eq}}+ \frac{\sqrt{t_{eq}t_{nr}}}{a_{eq}}\ln\left(\frac{t_{eq}}{t_{nr}}\right)+\frac{3\sqrt{t_{eq}t_{nr}}}{a_{eq}}\\
&\simeq& \frac{\sqrt{t_{eq}t_{nr}}}{a_{eq}}\left[5+\ln\left(\frac{t_{eq}}{t_{nr}}\right)\right].
\end{eqnarray}
where $a_{eq}$ is the scale factor at matter-radiation equality. Including entropy dilution the final expression is~ \cite{Merle:2013wta}
\begin{equation}
r_{\rm FS} \simeq \frac{\sqrt{t_{eq}t_{nr}}}{a_{eq}}\left[5+\ln\left(\frac{t_{eq}}{t_{nr}}\right)\right]/\xi^{1/3}.
\end{equation}
where the entropy dilution factor is given by
\begin{equation}
\xi=\frac{g_{\rm eff}({\rm high T})}{g_{\rm eff}({\rm t_0})}\simeq \frac{109.5}{3.36}\;.
\end{equation}
We have taken into account the scalar field $\sigma$ and the sterile neutrino $N$.
DM with a free-streaming scale larger than the size of a dwarf galaxy ($r_{FS}\gtrsim 0.1$ Mpc) constitutes HDM. The distinction between WDM and CDM is arbitrary. We follow ref.~\cite{Merle:2013wta} and consider DM with a free-streaming scale $r_{FS}< 0.01$ Mpc as CDM and DM with intermediate free-streaming scales as WDM.

The free-streaming horizon depends both on the mass of the scalar field and the mass of sterile neutrino. In \Figref{FreeStream} we plot the free-streaming horizon vs. the keV sterile neutrinos mass for different values of the scalar mass $m_\sigma=30,60,65,100,170,500$ GeV. The HDM (CDM) regions are coloured red (blue). Note that keV sterile neutrinos with $m_N=7.1$ keV are WDM for $m_\sigma\lesssim 170$ GeV and become too hot for larger frozen-in scalar masses. 
Demanding that the keV sterile neutrino accounts for the cosmologically observed DM abundance at $1\sigma$ and fixing  $\lambda_s=0.5$, we can plot the required value of the Higgs portal coupling $\lambda_{H\phi}$ as a function of the keV sterile neutrino mass $m_N$ and the frozen-in scalar mass $m_\sigma$. This is shown in \Figref{fig:scan}. The blue (red) coloured region indicates the CDM (HDM) region. The black lines are contour lines of equal $\lambda_{H\phi}$ assuming a vanishing active-sterile mixing angle, $\theta=0$. 
The jaggedness of the black contour lines appears because it is not possible to fix the DM abundance to a number, but we only demand it to lie within the $1\sigma$ allowed region. The grey shading in the background indicates the size of $\lambda_{H\phi}$. Darker regions correspond to larger values of $\lambda_{H\phi}$. The magenta line marks the Higgs mass $m_h$ and the orange line marks $m_h/2$. To the left of the orange line, Higgs decays will dominate the production of the frozen-in scalar $\sigma$ and the required value of $\lambda_{H\phi}$ is generically smaller. We find $\lambda_{H\phi}\lesssim 10^{-8}$ for $m_\sigma< m_h/2$ and $\lambda_{H\phi}\gtrsim 10^{-8}$ for $m_\sigma >m_h/2$. 
%
\section{UV Completion}\label{sec:UVcompletion}
So far we considered an effective operator to generate the neutrino mass. In this section we discuss one simple UV completion using the type-II seesaw mechanism. We introduce a SU(2) triplet
\begin{align}
\Delta&= \begin{pmatrix}\delta^+ & \delta^{++}\\ v_\Delta+\frac{1}{\sqrt{2}}(\delta^0+i\,a_\Delta) & -\delta^+\end{pmatrix} 
 \;,
\end{align}
which transforms under $Z_4$ as $\Delta\rightarrow -\Delta$. This introduces one more Yukawa coupling
\begin{equation}
-\mathcal{L}_\Delta =  \frac12 y_L L^T( i \sigma^2 \Delta) L +\hc\;,
\end{equation}
which leads to an additional contribution to neutrino mass after $\Delta$ acquires a vev, and the following additional terms in the scalar potential
\begin{align}\label{eq:Vdelta}
V_\Delta&= \mu_\Delta^2 \tr(\Delta^\dagger \Delta)+ \frac{\lambda_\Delta}{6} (\tr(\Delta^\dagger\Delta))^2
  + \frac{\lambda_\Delta^\prime}{12} \tr([\Delta^\dagger,\Delta]^2)\\\nonumber
&+\lambda_{H\Delta} H^\dagger H \tr(\Delta^\dagger\Delta)+ \lambda_{H\Delta}^\prime H^\dagger\left[\Delta^\dagger,\Delta\right]H
+\frac{\lambda_{\phi\Delta}}{2} \phi^2 \tr(\Delta^\dagger\Delta)\\\nonumber
&+ \left(\kappa H^T i\sigma_2 \Delta^\dagger H \phi +\hc\right)\;.
\end{align}
All parameters and vevs can be chosen real by rephasing $H$ and $\Delta$ and there is no spontaneous breaking of CP at tree-level. After $H$ and $\phi$ acquire vevs, a vev is induced for $\Delta$ given by
\begin{align}
v_\Delta&=\kappa\frac{3\sqrt{6}\mu_H^2\sqrt{\mu_\phi^2}}{\sqrt{\lambda_\phi}(3\lambda_{H\Delta}\mu_H^2+\lambda_H\mu_\Delta^2)}
\end{align}
in the limit of a weakly coupled field $\phi$ and a small vev $v_\Delta$. There are no charge-breaking minima. The non-vanishing Higgs triplet vev $v_\Delta$ will induce a contribution to the $\rho$ parameter. As the scalar $\phi$ and keV sterile neutrino are very weakly coupled, the constraints from the type-II seesaw model can be directly applied. See Ref.~\cite{Aoki:2012jj} for a recent analysis: $v_\Delta$ has to be less than 8 GeV, which follows from the tree level formula of the electroweak $\rho$ parameter. At leading order there is no mixing between the different scalar fields. The mixing is suppressed by the small Higgs portal couplings, $\lambda_{H\phi}, \lambda_{\Delta\phi}, \kappa$ and the assumption that the vev of $\Delta$ is induced via the coupling $\kappa$. 

In the scalar potential \eqref{eq:Vdelta} the only dimensionful coupling is $\mu_\Delta$.
For scales much smaller than $\mu_\Delta$ the type-II seesaw contribution is effectively described by the dimension 6 operator introduced in section \ref{section-Model}, i.e. we recover the previously discussed effective theory by integrating out $\Delta$.  
If the triplet $\Delta$ is heavy enough and its couplings to the scalar $\phi$ are sufficiently small\footnote{All quantum corrections to these couplings are either proportional to the coupling itself or the coupling $\lambda_{H\phi}$. 
Hence it is safe to assume that the couplings are sufficiently suppressed compared to $\lambda_{H\phi}$, such that they can be neglected.}, its contribution to the production of the scalar singlet $\sigma$ via triplet Higgs annihilations ($\Delta\Delta\leftrightarrow \sigma\sigma$) can be neglected: typical values of $\mu_\Delta \gtrsim 10$ TeV and $\lambda_{\phi\Delta}/\lambda_{H\phi}\lesssim\mathcal{O}(0.1)$ are sufficient.

Finally we want to comment on possible obstacles to a UV completion using the type-I seesaw mechanism~\cite{Minkowski:1977sc,Yanagida:1980,Glashow:1979vf,Gell-Mann:1980vs,Mohapatra:1980ia}. If the right-handed neutrinos carry the same $Z_4$ charge, they couple in the same way as the keV sterile neutrino. As they will be in thermal equilibrium via the Yukawa couplings to the lepton doublets, their couplings to $\phi$ and the keV sterile neutrino $N$ has to be sufficiently suppressed, in particular they are light compared to $v_\phi$ loosing the appeal of the type-I seesaw mechanism. However as we noted in section \ref{section-Model}, the $Z_4$ symmetry is not essential. The DM production mechanism works without the $Z_4$ symmetry, but the parameters have to be appropriately tuned to decouple the keV sterile neutrino and right-handed neutrinos in the usual way. See e.g. Ref.~\cite{Kusenko:2006rh,Petraki:2007gq}.

\section{Conclusion} \label{section-Conclusion}
We studied sterile neutrino dark matter production from the decay of a frozen-in scalar. The previous study of this production mechanism~\cite{Merle:2013wta} focused on heavy frozen-in scalars with masses above the Higgs mass and light keV sterile neutrinos with masses below $10$ keV turned out to have a free-streaming scale larger than the size of dwarf galaxies and are thus too hot and excluded. 

Motivated by the hints for an X-ray line at 3.55 keV, we are considering lighter frozen-in scalars. This leads to a smaller free-streaming horizon and thus lighter sterile neutrinos are allowed. We find that Higgs decay is the dominant production channel when kinematically allowed. For $m_\sigma>m_h/2$ Higgs decay becomes kinematically forbidden and the main production channels are annihilation into Higgs pairs as well as pairs of EW gauge bosons, particularly $W$-bosons. Above the EW phase transition the frozen-in scalar and thus the keV sterile neutrino is only produced by Higgs annihilations.

Furthermore we calculated the free-streaming horizon to show the viable region in parameter space, where sterile neutrinos are WDM or CDM. A 7.1 keV sterile neutrino requires the frozen-in scalar to be lighter than approximately 170 GeV.
Demanding that the keV sterile neutrino accounts for the cosmological DM abundance, we determined the necessary value of the Higgs portal coupling $\lambda_{H\phi}$. Above $m_h/2$ the coupling is $\lambda_{H\phi}\gtrsim 10^{-8}$, below $m_h/2$ the coupling is generically smaller, $\lambda_{H\phi}\lesssim 10^{-8}$.

Although we studied the DM production in an effective theory for neutrino mass, we showed a simple UV completion using the type-II seesaw mechanism~\cite{Magg:1980ut,Schechter:1980gr,Wetterich:1981bx,Lazarides:1980nt,Mohapatra:1980yp,Cheng:1980qt} in section \ref{sec:UVcompletion}. As long as the triplet is heavy enough and its couplings to the scalar $\phi$ are small enough, it can be neglected in the discussion of the production mechanism.

We discussed a minimal model of a keV sterile neutrino production from a frozen-in scalar. It might be interesting to promote the discrete $Z_4$ symmetry to a continuous $U(1)$. This introduces a (pseudo) Goldstone boson (pGB) which substantially modifies the production. As we expect the pGB to have strong couplings to the scalar $\sigma$ via its quartic interaction, it will be efficiently produced  and the cosmological DM abundance might be explained by a mixture of keV sterile neutrinos and the pGB.

\section*{Acknowledgements}
We would like to thank Alexander Merle and Viviana Niro for carefully reading our manuscript and their valuable comments. AA would like to thank Piyabut Burikham for his useful advice. This work has been supported by the Australian Research Council and Chulalongkorn University through Ratchadapisek Sompote Endowment Fund.

\appendix

\section{Cross sections and Decay Widths} \label{section-Cross}
The relevant cross sections $W_{ab} = 4 E_a E_b \sigma v$ for annihilation into fermions $W_{ff}$, vector bosons $W_{VV}$ and the Higgs $W_{hh}$\footnote{For simplicity we approximated the Higgs annihilation cross section and neglect higher order terms in $\lambda_{H\phi}$. They are shown in ref.~\cite{Guo:2010hq}.} are given by~\cite{Guo:2010hq}
\begin{align}
W_{hh}&=\frac{\lambda_{H\phi}^2}{16\pi} \sqrt{1-\frac{4m_h^2}{s}}\left(\frac{s+2m_h^2}{s-m_h^2}\right)^2\\
W_{t\bar t}&= \frac{\lambda_{H\phi}^2 m_t^2}{4\pi} \frac{1}{(s-m_h^2)^2+m_h^2\Gamma_h^2} \frac{(s-4m_t^2)^{3/2}}{\sqrt{s}}\\
W_{VV}&=\frac{\lambda_{H\phi}^2}{16\pi} \frac{1}{(s-m_h^2)^2+m_h^2\Gamma_h^2} \sqrt{1-\frac{4m_V^2}{s}} \left(1-\frac{4m_V^2}{s}+\frac{12m_V^4}{s^2}\right)\;.
\end{align}
The annihilation into $Z$ ($W$)-bosons is
\begin{align}
W_{ZZ}&=W_{VV}(m_V=m_Z)&
W_{WW}&=2\,W_{VV}(m_V=m_W)\;.
\end{align}

The partial Higgs decay width to scalars is given by
\begin{align}
\Gamma(h\rightarrow \sigma \sigma) =& \frac{|\lambda_{H\phi}|^2 v^2}{16\pi\, m_h}\left(1-\frac{4m_\sigma^2}{m_h^2}\right)^{1/2}
\end{align}
and the partial decay widths of $\sigma$ as well as the Higgs to two keV neutrinos $N$ are described by
\begin{align}
\Gamma(\sigma \rightarrow NN) =&\frac{y_N^2}{16\pi} m_\phi \left(1-\frac{4 m_N^2}{m_\sigma^2}\right)^{3/2}\\
\Gamma(h \rightarrow NN) =& \frac{y_N^2 \sin^2\gamma}{16\pi} m_h \left(1-\frac{4 m_N^2}{m_h^2}\right)^{3/2}\;,
\end{align}
where $y_N$ is chosen to be real without loss of generality and $\sin\gamma$ denotes the mixing between the Higgs and the scalar $\sigma$, which is defined in~\Eqref{Higgs-S Mixing}.

\section{Thermal Averages} \label{section-Thermal}
The thermal average of a partial decay width $\Gamma(X\to ii)$ of a particle $X$ with mass $m_X$ can be calculated as follows
\begin{align}
\langle\Gamma(X\to ii)  \rangle =&  \Gamma(X\to ii)  \frac{K_1(x)}{K_2(x)}
\end{align}
with $x=m_X/T$.
Following ref.~\cite{Guo:2010hq,Adulpravitchai:2011ei}, we write the thermally averaged annihilation cross section, $\langle \sigma v \rangle$, as 
\begin{equation}
\langle\sigma v(\sigma\sigma \to ab)\rangle = g_a g_b \frac{m_\sigma}{64\pi^4 x n_{eq}^2} \int_{4 m_\sigma^2}^\infty  W_{ab} \sqrt{1-\frac{4m_\sigma^2}{s}} \sqrt{s}K_1\left(\frac{x\sqrt{s}}{m_\sigma}\right)\mathrm{d}s
\end{equation}
where $x=m_\sigma/T$ and $n_{eq}$ denotes the equilibrium number density of $\sigma$. It is given by
\begin{equation}
n_{eq}=\frac{g_\sigma}{2\pi^2}\frac{m_\sigma^3}{x} K_2(x)
\end{equation}
where $g_\sigma=2 s_\sigma +1$ are the spin degrees of freedom of $\sigma$ and  $K_2(x)$ denotes the modified Bessel function of second kind.
$W_{ab}$ is defined as $4 E_a E_b \sigma v$. The relevant cross sections $W_{ab}$ are given in App.~\ref{section-Cross}.

\section{Effective Degree of Freedom} \label{section-effective}
We follow the discussion in \cite{Gondolo:1990dk}. The energy density and the entropy density are defined as
\begin{align}
\rho(T)=&g_{\rm eff}(T) \frac{\pi^2}{30} T^4 &
s(T) =& h_{\rm eff}(T) \frac{2 \pi^2}{45} T^3\;.
\end{align} 
In order to define the total effective number of degrees of freedom, we first have to define the effective number of degrees of freedom for energy and entropy for each individual particle species $i$
\begin{eqnarray}
g_i(T)&=&\frac{30}{\pi^2 T^4} \rho_i(T) = \frac{15 g_i}{\pi^4} x_i^4 \int_1^{\infty} \frac{y (y^2-1)^{1/2}}{e^{x_i y}+\eta_i} y dy \\
h_i(T)&=&\frac{45}{2 \pi^2 T^3} s_i(T) = \frac{45 g_i}{4\pi^4} x_i^4 \int_1^{\infty} \frac{y (y^2-1)^{1/2}}{e^{x_i y}+\eta_i} \frac{4y^2-1}{3y}dy \;, 
\end{eqnarray}
where $\rho_i(T)$ ($s_i(T)$) is the energy (entropy) density of each particle species and $x_i=m_i/T$ with $m_i$ being the mass of the particle species. 

The total energy effective degrees of freedom is
\begin{eqnarray}
g_{\rm eff}(T) &=& \sum\limits_{T_{d_i}<T}g_i(T) + \sum_{i \in {\rm dec}} g_i(T) \frac{T_i^4}{T^4} \equiv g_c(T) + \sum_{i \in {\rm dec}} g_i(T) \frac{T_i^4}{T^4} \;,
\end{eqnarray}
where $g_c(T)$ is the energy effective degree of freedom including all coupled species at temperature $T$. The second contribution includes all species that are already decoupled at temperature $T$. Like in \cite{Gondolo:1990dk}, we also choose $T_{d_i}=m_i/20$ for the decoupling temperature.

The total entropy effective degrees of freedom can be calculated as
\begin{eqnarray}
h_{\rm eff}(T) &=& h_c(T) \prod\limits_{i \in {\rm dec}}\Big(1+\frac{h_i(T_{d_i})}{h_c(T_{d_i})}\Big)\; ,
\end{eqnarray}
where $h_c(T)=\sum\limits_{T_{d_i}<T}h_i(T)$ is the entropy effective degree of freedom including all coupled species at temperature $T$ and the product here extends to all the species that are already decoupled at temperature $T$.
The effective degrees of freedom parameter $g_*^{1/2}$ is defined following refs.~\cite{Gondolo:1990dk, Merle:2013wta} as
\begin{equation}
\sqrt{g_*(T)}=\frac{h_{eff}(T)}{\sqrt{g_{eff}(T)}}\left(1+\frac13\frac{T}{h_{eff}(T)}\frac{\mathrm{d} h_{eff}(T)}{\mathrm{d}T}\right)\;.
\end{equation}

\newcommand{\eprint}[1]{arXiv: \href{http://arxiv.org/abs/#1}{\texttt{#1}}}

\bibliography{keV}

\end{document}